\documentclass[aps,twocolumn,superscriptaddress,citeautoscript,showpacs,amsmath,amssymb,amsfonts,floatfix,prl]{revtex4}
\usepackage{graphicx}
\usepackage{dcolumn}
\usepackage{bm}
\usepackage{times}
\usepackage{color}

\begin{document}

\title{Antiferromagnetic order in CaK(Fe$_{1-x}$Ni$_x$)$_4$As$_4$ and its interplay with superconductivity}

\author{A. Kreyssig}
\affiliation{Ames Laboratory, U. S. DOE, Ames, Iowa 50011, USA}
\affiliation{Department of Physics and Astronomy, Iowa State
University, Ames, Iowa 50011, USA}

\author{J. M. Wilde}
\affiliation{Ames Laboratory, U. S. DOE, Ames, Iowa 50011, USA}
\affiliation{Department of Physics and Astronomy, Iowa State
University, Ames, Iowa 50011, USA}

\author{A. E. B\"{o}hmer}
\affiliation{Ames Laboratory, U. S. DOE, Ames, Iowa 50011, USA}
\affiliation{Department of Physics and Astronomy, Iowa State
University, Ames, Iowa 50011, USA}

\author{W. Tian}
\affiliation{Neutron Scattering Division, Oak Ridge National 
Laboratory, Oak Ridge, Tennessee 37831, USA}

\author{W. R. Meier}
\affiliation{Ames Laboratory, U. S. DOE, Ames, Iowa 50011, USA}
\affiliation{Department of Physics and Astronomy, Iowa State
University, Ames, Iowa 50011, USA}

\author{Bing Li}
\affiliation{Ames Laboratory, U. S. DOE, Ames, Iowa 50011, USA}
\affiliation{Department of Physics and Astronomy, Iowa State
University, Ames, Iowa 50011, USA}

\author{B. G. Ueland}
\affiliation{Ames Laboratory, U. S. DOE, Ames, Iowa 50011, USA}
\affiliation{Department of Physics and Astronomy, Iowa State
University, Ames, Iowa 50011, USA}

\author{Mingyu Xu}
\affiliation{Ames Laboratory, U. S. DOE, Ames, Iowa 50011, USA}
\affiliation{Department of Physics and Astronomy, Iowa State
University, Ames, Iowa 50011, USA}

\author{S. L. Bud'ko}
\affiliation{Ames Laboratory, U. S. DOE, Ames, Iowa 50011, USA}
\affiliation{Department of Physics and Astronomy, Iowa State
University, Ames, Iowa 50011, USA}

\author{P. C. Canfield}
\affiliation{Ames Laboratory, U. S. DOE, Ames, Iowa 50011, USA}
\affiliation{Department of Physics and Astronomy, Iowa State
University, Ames, Iowa 50011, USA}

\author{R. J. McQueeney}
\affiliation{Ames Laboratory, U. S. DOE, Ames, Iowa 50011, USA}
\affiliation{Department of Physics and Astronomy, Iowa State
University, Ames, Iowa 50011, USA}

\author{A. I. Goldman}
\affiliation{Ames Laboratory, U. S. DOE, Ames, Iowa 50011, USA}
\affiliation{Department of Physics and Astronomy, Iowa State
University, Ames, Iowa 50011, USA}

\date{\today}

\begin{abstract}
The magnetic order in CaK(Fe$_{1-x}$Ni$_x$)$_4$As$_4$ (1144) 
single crystals ($x$\,=\,0.051 and 0.033) has been studied by 
neutron diffraction.  We observe magnetic Bragg peaks associated 
to the same propagation vectors as found for the collinear stripe 
antiferromagnetic (AFM) order in the related BaFe$_2$As$_2$ (122) 
compound.  The AFM state in 1144 preserves tetragonal symmetry 
and only a commensurate, non-collinear structure with a hedgehog 
spin-vortex crystal (SVC) arrangement in the Fe plane and simple 
AFM stacking along the $\boldsymbol{c}$ direction is consistent 
with our observations.  The SVC order is promoted by the reduced 
symmetry in the FeAs layer in the 1144 structure.  The long-range 
SVC order coexists with superconductivity, however, similar to 
the doped 122 compounds, the ordered magnetic moment is gradually 
suppressed with the developing superconducting order parameter.  
This supports the notion that both collinear and non-collinear 
magnetism and superconductivity are competing for the same 
electrons coupled by Fermi surface nesting in iron arsenide 
superconductors. 
\end{abstract}


\maketitle

\begin{figure}
\centering\includegraphics[width=1.00\linewidth]{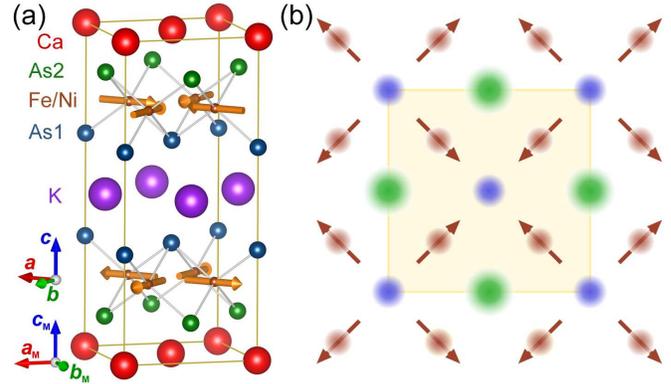} 
\caption{\label{fig:fig1}(Color online) Chemical and 
antiferromagnetic structure of CaK(Fe$_{1-x}$Ni$_x$)$_4$As$_4$. 
(a) Antiferromagnetic tetragonal unit cell with 
$\boldsymbol{a_{\rm{M}}}$, $\boldsymbol{b_{\rm{M}}}$, and 
$\boldsymbol{c_{\rm{M}}}$ which is doubled and 45$^{\circ}$ 
rotated in the ($\boldsymbol{ab}$) plane with respect to the 
chemical unit cell with $\boldsymbol{a}$, $\boldsymbol{b}$, and 
$\boldsymbol{c}$.  The arrows represent the antiferromagnetically 
ordered Fe moments.  (b) Arrangement of the magnetic Fe moments 
in a FeAs layer.}
\end{figure}

The diversity of iron-based superconductors has provided many 
insights into the relationships between their structure, 
magnetism and superconductivity.  Iyo \emph{et 
al.}\cite{Iyo_2016} opened a new avenue of research with the 
discovery of the $AeA$Fe$_4$As$_4$ ($Ae$\,=\,Ca,\,Sr; 
$A$\,=\,K,\,Rb,\,Cs) (1144) compounds.  Although closely related 
to the much studied $Ae$Fe$_2$As$_2$ (122) 
system\cite{Canfield_2010, Paglione_2010}, there are important 
differences in their structure and symmetry.  For example, the 
cation planes in CaKFe$_4$As$_4$ alternate between Ca and K as 
illustrated in Fig.\,\ref{fig:fig1}.  In consequence, there are 
two distinct As sites, As1 and As2, neighboring K and Ca, 
respectively, rather than one As site found in CaFe$_2$As$_2$ and 
KFe$_2$As$_2$.  The local symmetry at the Fe sites is reduced 
from tetragonal to orthorhombic.\cite{Iyo_2016}  The space group 
for CaKFe$_4$As$_4$ is primitive tetragonal, 
\textit{P}\,4/\textit{m\,m\,m}, rather than body-centered 
tetragonal, \textit{I}\,4/\textit{m\,m\,m}, for CaFe$_2$As$_2$ 
and KFe$_2$As$_2$.

CaKFe$_4$As$_4$ shows bulk superconductivity below 
$T_{\rm{c}}=$\,35\,K\cite{Iyo_2016, Meier_2016}.  KFe$_2$As$_2$ 
is also superconducting but with low $T_{\rm{c}}\sim$\,3.8\,K in 
comparison\cite{Sasmal_2008}.  In contrast, CaFe$_2$As$_2$ is not 
superconducting at ambient pressure and requires chemical 
substitution to realize superconductivity, e.\,g. electron-doping 
by partially replacing Fe with Co or Ni or hole-doping by 
substituting K for Ca\cite{Ran_2012, Ran_2014, Wang_2013}.  From 
this perspective of electron count, stoichiometric 
CaKFe$_4$As$_4$ may be viewed as nearly optimally hole-doped 
CaFe$_2$As$_2$, but without disorder arising from Ca and K 
randomly occupying the same site.

Partial substitution of Co or Ni for Fe in CaKFe$_4$As$_4$ 
(electron doping) should, in principle, shift the ground state 
from superconducting to antiferromagnetically (AFM) 
ordered\cite{Zinth_2011}.  Indeed, superconductivity is 
suppressed and signatures of an additional phase transition have 
been observed in electric resistance and specific heat 
measurements\cite{Meier_2018}.  $^{57}$Fe M\"{o}ssbauer studies 
have identified this additional phase transition as magnetic in 
nature\cite{Meier_2018}.  However, the orthorhombic lattice 
distortion that accompanies AFM in CaFe$_2$As$_2$ was not 
observed in CaK(Fe$_{1-x}$Ni$_x$)$_4$As$_4$\cite{Meier_2018}.  
Furthermore, $^{75}$As Nuclear Magnetic Resonance (NMR) studies, 
together with symmetry analysis have proposed that the AFM order 
of the Fe moments is hedgehog spin-vortex crystal (SVC) order in 
the Fe planes as shown in Fig.\,\ref{fig:fig1}(b). This order is 
characterized by non-collinear Fe moments featuring an 
alternating all-in and all-out motif around the As1 
sites\cite{Meier_2018}.  The temperature dependence of the 
nuclear spin-lattice relaxation rate provides evidence that the 
AFM order coexists microscopically with 
superconductivity\cite{Ding_2017}.

Many open questions remain regarding the magnetic ordering in 
CaK(Fe$_{1-x}$Ni$_x$)$_4$As$_4$:  What is the spatial extent -- 
long-range order or short-range correlations?  Is the magnetic 
order commensurate or incommensurate with the lattice?  What is 
the nature of the magnetic correlations along the 
$\boldsymbol{c}$ direction -- AFM or ferromagnetic (FM)?  Is 
there an interplay between magnetism and superconductivity?  Here 
we address the preceding questions via neutron diffraction 
measurements.

In this communication, we describe a neutron diffraction study of the 
magnetic order in electron-doped CaK(Fe$_{1-x}$Ni$_x$)$_4$As$_4$ 
single crystals with $x$\,=\,0.051 and 0.033.  In both samples, 
the Fe magnetic moments order antiferromagnetically in a 
long-range, commensurate and non-collinear structure with a 
hedgehog spin-vortex crystal arrangement in the Fe planes and 
simple AFM stacking along the $\boldsymbol{c}$ direction.  This 
magnetic order preserves the tetragonal symmetry and coexists 
with the superconductivity below $T_{\rm{c}}$.  For 
$x$\,=\,0.033, the ordered magnetic moment is gradually 
suppressed below $T_{\rm{c}}$.  This is similar to the behavior 
observed for electron-doped Ba(Fe$_{1-x}$$M_x$)$_2$As$_2$ with 
$M$\,=\,Co, Ni, or Rh\cite{Fernandes_2010, Kreyssig_2010, 
Luo_2012} and hole-doped 
BaK$_{1-x}$Fe$_2$As$_2$\cite{Munevar_2013} but contrasts with the 
mutual exclusion of AFM and superconductivity in electron-doped 
Ca(Fe$_{1-x}$$M_x$)$_2$As$_2$\cite{Ran_2012, Ran_2014, 
Sapkota_2018}.  

Single crystals of CaK(Fe$_{1-x}$Ni$_x$)$_4$As$_4$ with 
$x$\,=\,0.051(1) and $x$\,=\,0.033(1) and masses of 4.3(1)\,mg 
and 3.7(1)\,mg, respectively, were grown from a high-temperature 
transition-metal arsenic solution as described in 
Refs.\,[\onlinecite{Meier_2017, Meier_2018}].  Composition was 
determined via wavelength-dispersive x-ray spectroscopy employing 
a JEOL JXA-8200 microprobe system on cleaved surfaces of crystals 
from the same batches\cite{Meier_2017}.  No deviation of the 
Ni-concentrations outside of the given statistical error are 
observed for either batch.  The AFM transition temperatures 
$T_{\rm{N}}$\,=\,50.6(5)\,K and 42.9(5)\,K for $x$\,=\,0.051 and 
0.033, respectively, are inferred from temperature-dependent 
electrical-resistance and heat-capacity measurements using a 
Janis Research SHI-950T 4\,Kelvin closed-cycle refrigerator and a 
Quantum Design (QD), Physical Property Measurement Systems.  
Employing a QD, Magnetic Property Measurement System, no 
signatures of impurity phases were observed in magnetization 
measurements on the specific samples used in this study and 
$T_{\rm{c}}$ was determined to be 9.0(8)\,K and 21.0(4)\,K for 
$x$\,=\,0.051 and 0.033, respectively.  High-energy x-ray 
diffraction measurements were performed similar to those 
described in Ref.\,[\onlinecite{Meier_2018}] on samples from the 
same batches and demonstrated that single crystals of both Ni 
concentrations maintain the same tetragonal crystallographic 
structure down to temperatures of 7\,K.

Neutron diffraction measurements were performed on the HB--1A 
FIE--TAX triple-axis spectrometer at the High Flux Isotope 
Reactor, Oak Ridge National Laboratory, using a fixed incident 
energy of 14.6\,meV, and effective collimations of 
40$^{\prime}$\,-\,40$^{\prime}$\,-\,S\,-\,40$^{\prime}$\,-\,80$^{\prime}$ 
in front of the pyrolytic graphite (PG) monochromator, between 
the monochomator and sample, between the sample and PG analyzer, 
and between the analyzer and detector, respectively.  Two PG 
filters were used to minimize contamination from higher 
harmonics.  The samples were mounted in a helium-filled aluminum 
can attached to the cold finger of a helium closed-cycle 
refrigerator with the ($H$\,$H$\,$L$) plane coincident with the 
scattering plane of the instrument.  Both samples exhibited 
resolution-limited rocking scans indicating high-quality single 
crystals. 

\begin{figure}
\centering\includegraphics[width=0.95\linewidth]{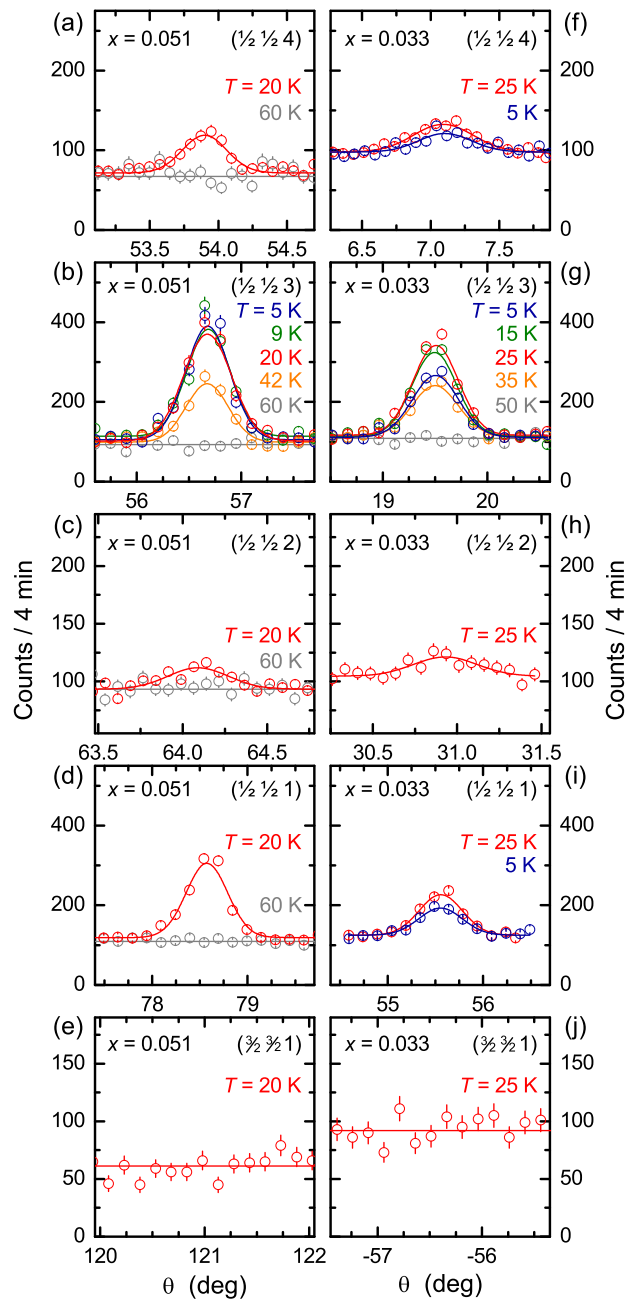} 
\caption{\label{fig:fig2}(Color online) Magnetic Bragg peaks of 
CaK(Fe$_{1-x}$Ni$_x$)$_4$As$_4$ measured by neutron diffraction 
rocking scans on single crystals with (a)-(e) $x$~=~0.051 and 
(f)-(j) $x$~=~0.033 at selected temperatures. The data are 
normalized to a monitor value of 240\,mcu (monitor count units) 
which corresponds to 4 min of counting time.}
\end{figure}

\begin{figure}
\centering\includegraphics[width=0.95\linewidth]{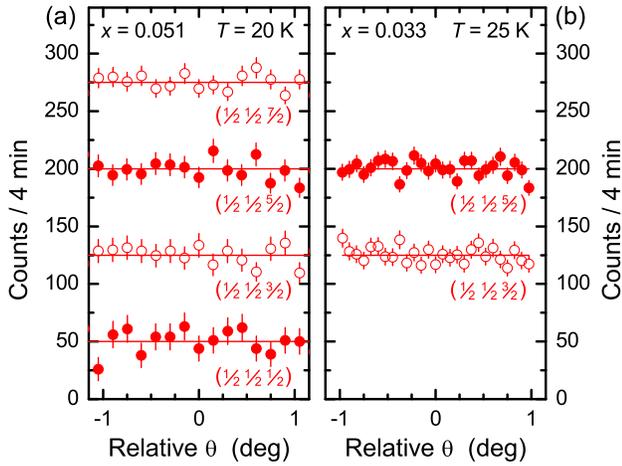} 
\caption{\label{fig:fig3}(Color online) Neutron diffraction 
rocking scans on CaK(Fe$_{1-x}$Ni$_x$)$_4$As$_4$ single crystals 
with (a) $x$~=~0.051 and (b) $x$~=~0.033 measured at AFM Bragg 
peak positions with half integer $L$ associated with a unit cell 
doubling along $\boldsymbol{c}$, and at a temperature well below 
$T_{\rm{N}}$ and slightly above $T_{\rm{c}}$. The data are 
normalized to a monitor value of 240\,mcu and are offset for 
clarity.}
\end{figure}

\begin{figure}
\centering\includegraphics[width=0.95\linewidth]{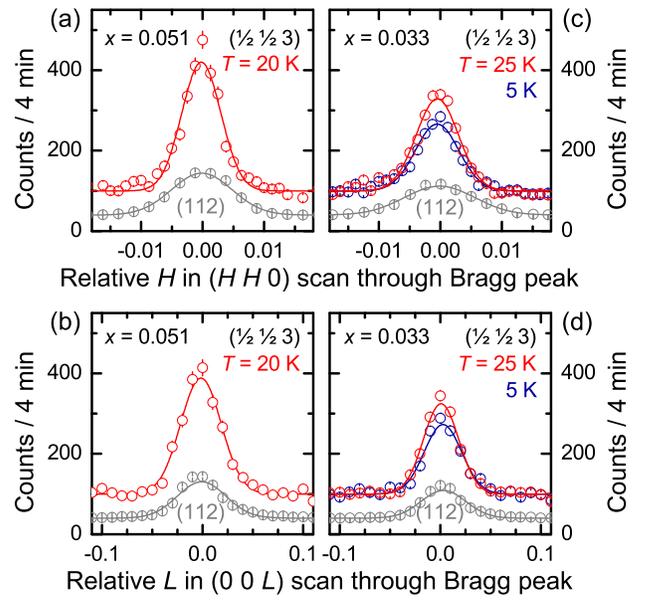} 
\caption{\label{fig:fig4}(Color online) Neutron diffraction scans 
along ($H$\,$H$\,0) and (0\,0\,$L$) through the 
($\frac{1}{2}$\,$\frac{1}{2}$\,3) AFM Bragg peak for  
CaK(Fe$_{1-x}$Ni$_x$)$_4$As$_4$ single crystals with (a), (b) 
$x$\,=\,0.051 and (c), (d) $x$\,=\,0.033 and at selected 
temperatures.  The data are normalized to a monitor value of 
240\,mcu.  Similar scans through the (1\,1\,2) Bragg peak 
characterizing the chemical structure and the resolution 
conditions are shown for comparison with the intensity divided by 
a factor of 2,000.}
\end{figure}

Magnetic Bragg peaks at positions 
($\frac{1}{2}$\,$\frac{1}{2}$\,$L$) with integer $L$ develop 
below the N\'eel temperature $T_{\rm{N}}$ as shown in 
Fig.\,\ref{fig:fig2}.  These Bragg peaks are consistent with AFM 
order characterized by a doubling, and 45$^{\circ}$ rotation, of 
the magnetic unit cell in the ($\boldsymbol{ab}$) plane with 
respect to the chemical unit cell.  Magnetic Bragg peaks at 
($\frac{1}{2}$\,$\frac{1}{2}$\,$L$) with half integer $L$ are 
absent, as shown in Fig.\,\ref{fig:fig3}, signaling that the 
magnetic and chemical unit cells have same lengths along 
$\boldsymbol{c}$.  Rocking scans through the AFM Bragg peaks 
displayed in Fig.\,\ref{fig:fig2} show the same shape and widths 
at all measured temperatures, as do scans through the 
($\frac{1}{2}$\,$\frac{1}{2}$\,3) AFM and the (1\,1\,2) nuclear 
Bragg peaks along the ($H$\,$H$\,0) and (0\,0\,$L$) directions, 
as presented in Fig.\,\ref{fig:fig4}.  Taken together, these data 
demonstrate that the AFM Bragg peaks are resolution limited, 
which places a lower limit on the AFM correlation length of 
$\sim$60\,nm, and show that the AFM order is commensurate.

At first glance, the appearance of AFM Bragg peaks at 
($H$\,$H$\,$L$) with half integer $H$ and integer $L$ might be 
attributed to the ubiquitous stripe-like AFM in many other iron 
arsenides\cite{Paglione_2010, Lumsden_2010, Dai_2015, 
Goldman_2008}.  However, our evidence of a tetragonal unit cell 
in the AFM ordered phase is inconsistent with the orthorhombic 
distortion intrinsically linked to the stripe-like 
AFM\cite{Kim_2010, Wasser_2015, Allred_2016, Nandi_2010, 
Kreyssig_2010, Goldman_2008}.  Alternatively, tetragonal AFM 
order can be constructed by coherent superposition of the two 
orientations of stripe-like modulations\cite{Kim_2010, 
Wasser_2015, Allred_2016, Cvetkovic_2013, Fernandes_2014, 
Holloran_2017}.  These orientations arise from the pair of 
symmetry-equivalent propagation vectors 
$\boldsymbol{\tau}_{1}$\,=\,($\pi$,\,0) and 
$\boldsymbol{\tau}_{2}$\,=\,(0,\,$\pi$) in units of the Brillouin 
zone of the Fe square lattice.

Three different AFM arrangements of the Fe moments in the 
($\boldsymbol{ab}$) plane are distinguished by the relative 
orientation of the AFM ordered Fe-moment components 
$\boldsymbol{\mu}_{\textit{i}}$ to their corresponding 
propagation vectors $\boldsymbol{\tau}_{\textit{i}}$: (i) 
$\boldsymbol{\mu}_{\textit{i}}$ in the ($\boldsymbol{ab}$) plane 
and parallel to $\boldsymbol{\tau}_{\textit{i}}$, (ii) 
$\boldsymbol{\mu}_{\textit{i}}$ in the ($\boldsymbol{ab}$) plane 
and perpendicular to $\boldsymbol{\tau}_{\textit{i}}$, and (iii) 
$\boldsymbol{\mu}_{\textit{i}}$ along the $\boldsymbol{c}$ 
direction\cite{Cvetkovic_2013, Christensen_2015, Fernandes_2016, 
Holloran_2017, Meier_2018}.  These AFM structures have been 
described as (i) hedgehog SVC order, (ii) loops SVC order, and 
(iii) spin charge-density wave (SCDW) order\cite{Cvetkovic_2013, 
Wang_2015, Allred_2016} as illustrated in Fig.\,1 of 
Ref.\,[\onlinecite{Meier_2018}].  In each case, the AFM ordered 
Fe planes can be either AFM or FM stacked along 
$\boldsymbol{c}$.  

Table\,\ref{tab:tab1} compares the measured integrated 
intensities of selected AFM Bragg peaks to their intensities 
calculated using Fullprof\cite{Rodriguez_1993} for each of these 
six cases.  Qualitative comparison of intensities between the 
different columns yields the result that only the hedgehog SVC 
order with AFM stacking along the $\boldsymbol{c}$ direction is 
consistent with the observations for both samples, e.\,g. the 
($\frac{1}{2}\,\frac{1}{2}\,3$) Bragg peaks are the strongest and 
the ($\frac{3}{2}\,\frac{3}{2}\,1$) Bragg peaks are very weak for 
both samples and this AFM order.  This hedgehog SVC order is 
consistent with the arrangement of the Fe moments in the 
($\boldsymbol{ab}$) plane proposed by Meier \textit{et 
al.}\cite{Meier_2018} and is illustrated in 
Fig.\,\ref{fig:fig1}.  The magnetic space group is 
\textit{P}$_{C}$\,4/\textit{m\,b\,m} (BNS) with respect to the 
AFM unit cell\cite{Belov_1955}, and 
\textit{P}$_{P}$\,4'/\textit{m\,m\,m'} (OG) with respect to the 
chemical unit cell\cite{Opechowski_1965}.  The AFM order can be 
described as a two--$\boldsymbol{\tau}$ structure with 
propagation vectors $\boldsymbol{\tau}_{1}$\,=\,($\pi$,\,0) and 
$\boldsymbol{\tau}_{2}$\,=\,(0,\,$\pi$), or 
$\boldsymbol{\tau}_{1}$\,=\,($\frac{1}{2}$\,$\frac{1}{2}$\,1) and 
$\boldsymbol{\tau}_{2}$\,=\,($\overline{\frac{1}{2}}$\,$\frac{1}{2}$\,1) 
in reciprocal lattice units, modulating Fe moments 
{$\boldsymbol{\mu}_{\textit{i}}$ in the 
($\boldsymbol{ab}$)\,plane with 
{$\boldsymbol{\mu}_{\textit{i}}$\,$\parallel$\,$\boldsymbol{\tau}_{\textit{i}}$}.

From fitting the measured integrated intensities of the AFM Bragg 
peaks listed in Tab.\,\ref{tab:tab1} against the calculated 
values for this hedgehog SVC structure, the total AFM ordered 
moment per transition-metal site is determined as 
0.37(10)\,$\mu_{\rm{B}}$ and 0.34(10)\,$\mu_{\rm{B}}$ for the 
$x$\,=\,0.051 sample at $T$\,=\,20\,K and the $x$\,=\,0.033 
sample at $T$\,=\,25\,K, respectively.  The value for 
$x$\,=\,0.051 is in good agreement with the hyperfine field at 
the Fe position determined from $^{57}$Fe M\"{o}ssbauer 
measurements\cite{Meier_2018}.

\begin{table}
\caption{\label{tab:tab1} Integrated intensity of selected AFM 
Bragg peaks measured on both CaK(Fe$_{1-x}$Ni$_x$)$_4$As$_4$ 
single crystals and calculated for a total magnetic moment of 
0.37\,$\mu_{\rm{B}}$ per transition-metal site for the SVC 
orders, and alternating 0.74\,$\mu_{\rm{B}}$ and 
0\,$\mu_{\rm{B}}$ per transition-metal site for the SCDW order. 
The intensities are in arbitrary units and normalized to the 
intensities of ten selected chemical Bragg peaks.}
 \begin{tabular}{|c|c|c|c|c|c|c|c|c|}
 \hline
   & \multicolumn{2}{c|}{Measurement} & \multicolumn{6}{c|}{Calculation} \\
   & & & \multicolumn{2}{c|}{Hedgehog SVC}& \multicolumn{2}{c|}{Loops SVC} & \multicolumn{2}{c|}{SCDW} \\
   & $x$\,= & $x$\,= & \multicolumn{2}{c|}{in\,($\boldsymbol{ab}$)\,plane:} & \multicolumn{2}{c|}{in\,($\boldsymbol{ab}$)\,plane:} & \multicolumn{2}{c|}{$\boldsymbol{\mu}_{\textit{i}}$\,$\parallel$\,$\boldsymbol{c}$} \\
  AFM & \,0.033\, & \,0.051\, & \multicolumn{2}{c|}{$\boldsymbol{\mu}_{\textit{i}}$\,$\parallel$\,$\boldsymbol{\tau}_{\textit{i}}$} & \multicolumn{2}{c|}{$\boldsymbol{\mu}_{\textit{i}}$\,$\perp$\,$\boldsymbol{\tau}_{\textit{i}}$} & \multicolumn{2}{c|}{} \\
  Bragg & $T$\,= & $T$\,=  & \multicolumn{2}{c|}{along\,$\boldsymbol{c}$:} & \multicolumn{2}{c|}{along\,$\boldsymbol{c}$:} & \multicolumn{2}{c|}{along\,$\boldsymbol{c}$:} \\
  peak & 25\,K & 20\,K & ~~AFM~~ & FM & ~AFM~ & FM & AFM & \,FM \\
 \hline\hline
 ($\frac{1}{2}\,\frac{1}{2}\,4$) &  22    &  18    &  19 &  80 &  26 & 107 &   13 &  54 \\
 ($\frac{1}{2}\,\frac{1}{2}\,3$) & 143    & 144    & 130 &  17 & 207 &  26 &  155 &  20 \\
 ($\frac{1}{2}\,\frac{1}{2}\,2$) &  12    &   6    &   9 & 164 &  21 & 384 &   24 & 441 \\
 ($\frac{1}{2}\,\frac{1}{2}\,1$) &  69    &  95    &  99 &   1 & 634 &   8 & 1071 &  14 \\
 ($\frac{3}{2}\,\frac{3}{2}\,1$) & $<$\,2 & $<$\,2 &   1 & 0.1 &  32 & 0.4 &   63 &   1 \\
 \hline
 \end{tabular}
\end{table}
 
\begin{figure}
\centering\includegraphics[width=0.95\linewidth]{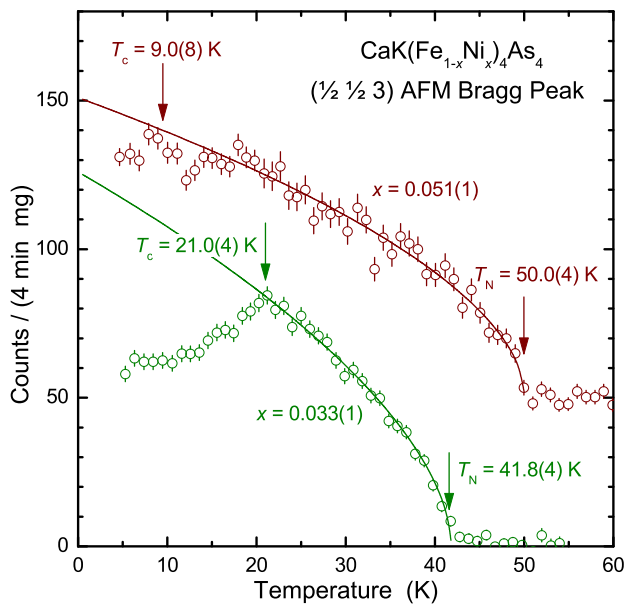} 
\caption{\label{fig:fig5}(Color online) Temperature dependence of 
the intensity of the ($\frac{1}{2}$\,$\frac{1}{2}$\,3) AFM Bragg 
peak for CaK(Fe$_{1-x}$Ni$_x$)$_4$As$_4$ single crystals with 
$x$~=~0.051(1) and $x$~=~0.033(1).  The measured counts represent 
the intensity because the widths of the AFM Bragg peaks don't 
change with temperature.  The data are offset for clarity and 
normalized to the mass of the sample and a monitor value of 
240\,mcu.  The lines represent power-law fits as described in the 
text.  The transition temperatures are marked by arrows at values 
for $T_{\rm{N}}$ determined from the fits and for $T_{\rm{c}}$ 
determined from magnetization measurements.}
\end{figure}

Figure\,\ref{fig:fig5} shows the temperature dependence of the 
intensity measured at the ($\frac{1}{2}$\,$\frac{1}{2}$\,3) AFM 
Bragg peak position for both samples which is proportional to the 
square of the AFM moment, the AFM order parameter.  The AFM 
ordering for both samples is well-described as a second-order 
phase transition by a power law with $T_{\rm{N}}$\,=\,50.0(4)\,K 
and 41.8(4)\,K for the $x$\,=\,0.051 and $x$\,=\,0.033 samples, 
respectively, in agreement with results from the transport and 
thermodynamic measurements described earlier.  The critical 
exponent for both samples is $\beta$\,=\,0.29(2) that is close to 
the value of 0.33 expected for a three-dimensional Heisenberg 
system.  This behavior is consistent with the fact that the 1144 
structure already features the necessary broken structural 
symmetry which allows for the onset of the hedgehog SVC order to 
be second order. 

As temperature is lowered below the superconducting transition 
temperature, $T_{\rm{c}}$, AFM order persists and coexists with 
superconductivity in CaK(Fe$_{1-x}$Ni$_x$)$_4$As$_4$.  For 
$x$~=~0.051, the AFM order parameter increases smoothly to the 
lowest temperature measured.  However, for $x$~=~0.033, the 
magnetic order parameter clearly decreases gradually below 
$T_{\rm{c}}$.  This is reminiscent of what has previously been 
observed for Ba(Fe$_{1-x}$$M_x$)$_2$As$_2$ with $M$\,=\,Co, Ni, 
and Rh\cite{Fernandes_2010, Luo_2012, Kreyssig_2010}, and 
Ba$_{1-x}$K$_x$Fe$_2$As$_2$\cite{Munevar_2013} and presents such 
behavior in a second family of iron-based superconductors with 
different AFM order.  

The 1144 compounds fill a unique and interesting niche in the 
family of iron-based superconductors due to the reduced symmetry 
in the FeAs layers.  In most iron-based superconductors, the Fe 
site has tetragonal symmetry and a high-symmetry direction can be 
found every 45$^{\circ}$ in the ($\boldsymbol{ab}$) plane, e.\,g. 
the $\boldsymbol{a}$, $\boldsymbol{b}$, and diagonal directions.  
In contrast, the Fe site has orthorhombic symmetry in the 1144 
compounds with high-symmetry directions only every 90$^{\circ}$.  
In this environment, the magneto-crystalline anisotropy and 
spin-orbit coupling will constrain the Fe magnetic moments to lie 
in these high-symmetry directions exemplified by the SVC motif in 
Fig. 1.  In contrast, if stripe-type AFM were occur, the Fe 
moments would lie along arbitrary directions. This leads to a 
preference for SVC orders in 1144 compounds\cite{Cvetkovic_2013, 
Fernandes_2014, Holloran_2017}.

In both, the 1144 and 122 compounds, the AFM orders are related 
to the same propagation vector ($\pi$,\,0) and the 
symmetry-equivalent (0,\,$\pi$) but the directions of the AFM 
ordered Fe moments are different.  In the 1144 compounds, the Fe 
moments are non-collinear arranged in the SVC motif and lie 
45$^{\circ}$ to those of the collinear stripe-like order in the 
122 system.  However, both AFM orders demonstrate a similar 
interplay with superconductivity.  This suggests that whereas 
their common underling propagation vectors may be important, the 
orientation of the ordered moments and their collinear or 
non-collinear arrangement apparently are not, and so e.\,g. 
scattering processes of Cooper pairs on magnetic moments, which 
would change significantly for different moment directions and 
non-/collinearity, seem not the dominating factor for the 
interplay between superconductivity and AFM.  Instead it points 
to superconductivity and magnetism competing for the same 
electrons coupled by the same wave vector, i.\,e. the Fermi 
surface nesting vector ($\pi$,\,0)\cite{Fernandes_2010}.  Hereby, 
the AFM order plays the role of an intrinsic Josephson coupling 
and provides a sensitive probe to the relative phase of the 
Cooper-pair wave functions\cite{Fernandes_2010}:  Whereas a 
pairing mechanism with $s^{++}$ symmetry is intrinsically 
unsuitable for coexistence of superconductivity with AFM, an 
$s^{+-}$ state may or may not coexist with AFM depending on 
details of the band structure.  The observed coexistence and 
competition with the gradual suppression of the ordered magnetic 
moment below $T_{\rm{c}}$ supports then strongly a pairing 
mechanism with $s^{+-}$ symmetry in the 1144 system consistent 
with the two-gap $s+s$ model deducted from a muon spectroscopy 
study\cite{Biswas_2017}, and as has previously been established 
for the 122 iron-arsenide superconductors\cite{Fernandes_2010, 
Fernandes_2014}.

Summarizing, we have shown via neutron diffraction measurements 
that the magnetic order in CaK(Fe$_{1-x}$Ni$_x$)$_4$As$_4$ is 
long-range and commensurate to the lattice.  The Fe moments order 
in a hedgehog SVC motif in each Fe plane and are AFM stacked 
along the $\boldsymbol{c}$ direction.  The 1144 compounds are 
unique in the family of iron-based superconductors due to reduced 
symmetry in the FeAs layers promoting SVC order.  This 
non-collinear AFM order coexists with superconductivity, however, 
the magnetic order parameter decreases gradually below 
$T_{\rm{c}}$, reminiscent of what has previously been observed 
for collinear stripe-like AFM in 122 compounds.

We are grateful for excellent assistance by D.\,S.\,Robinson with 
performing the high-energy x-ray diffraction experiments and for 
helpful discussions with P.\,P.\,Orth.  Work at the Ames 
Laboratory was supported by the US Department of Energy (DOE), 
Basic Energy Sciences, Division of Materials Sciences and 
Engineering, under Contract No. DEAC02-07CH11358.  A portion of 
this research used resources at the High Flux Isotope Reactor, a 
US DOE Office of Science User Facility operated by the Oak Ridge 
National Laboratory.  This research used resources of the 
Advanced Photon Source, a US DOE Office of Science User Facility 
operated for the US DOE Office of Science by Argonne National 
Laboratory under Contract No. DE-AC02-06CH11357.  W.\,R.\,Meier 
was supported by the Gordon and Betty Moore Foundation's EPiQS 
Initiative through Grant GBMF4411.

\bibliographystyle{apsrev}
\bibliography{CaKFeNi4As4_AFMSC}

\end{document}